\documentclass[jkps,preprint,fleqn,showpacs,showkeys]{revtex4}
\usepackage{graphicx}
\usepackage{amssymb}
\usepackage{amsmath}
\usepackage{epstopdf}
\usepackage{bm}
\begin{document}
\setcounter{page}{1}
\title{Effect of optimal uncoupling in enhancing synchronization stability in coupled chaotic systems}
\author{G. Sivaganesh}
\affiliation{Department of Physics, Alagappa Chettiar College of Engineering $\&$ Technology, Karaikudi, Tamilnadu-630 004, India}
\author{B. D. Sharmila}
\affiliation{Government Girls Higher Secondary School, Muthupattanam, Karaikudi, Tamilnadu-630 001, India}
\author{A. Arulgnanam}
\email{gospelin@gmail.com}
\affiliation{Department of Physics, St. John's College, Palayamkottai, Tamilnadu-627 002, India}


\begin{abstract}
In this paper, we report a novel approach for studying the effect of optimal uncoupling on the stability of synchronization in coupled chaotic systems. The clipping of phase space of the driven system having an orientation along the coordinate axes revealing the nature of coupling of the state variables of coupled systems is identified in certain coupled third-order chaotic systems. The stability of synchronization is studied through the {\emph{Master Stability Function}} (MSF). The optimal directions of implementing the clipping width to achieve stable synchronization is observed by studying the effectiveness of clipping fraction and the sufficient range of orientation to identify the optimal directions is reported. The functional work steps for identifying the optimal directions are presented and the synchronization of the response system with the drive within the clipped region of phase space for different orientations of clipping width are studied. The stability of synchronization for different orientations of clipping widths and the two-parameter bifurcation diagram indicating the negative valued MSF regions obtained for the optimal direction of clipping width are presented. The application of the method of optimal uncoupling in identifying the direction of implication of clipping width is discussed and the range of orientation over which the clipping width has to be varied is generalized.
\end{abstract}

\pacs{05.45.Xt, 05.45.-a}

\keywords{coupled oscillators; synchronization; optimal uncoupling; master stability function}

\maketitle

\section{INTRODUCTION}

The dynamical process of synchronization in coupled chaotic systems has greatly influenced researchers because of the sensitive dependence of chaos on initial conditions \cite{Pecora1990,Pikovsky2003}. The high complexity and unpredictability prevailing in the dynamics of chaotic systems requires a complete understanding on the synchronization dynamics of coupled systems as it has potential applications in secure transmission of information signals  \cite{Rulkov1992,Chua1992,Chua1993,Murali1993,Boccaletti2002,Chen2020}. Several higher and low-dimensional chaotic systems have been studied for synchronization and numerous electronic circuit systems have been analyzed for the application of chaos synchronization to secure communication \cite{Chua1992,Murali1993,Oppenheim1992,Murali1994,Murali1997,Koronovskii2009,Wu2019,Wang2019,Wang2019a}. The important requirement for signal transmission by chaos synchronization is that the coupled systems must exist in stable synchronized states over greater values of coupling strength. The existence of coupled chaotic systems in stable synchronized states is observed through the evaluation of the {\emph{Master Stability Function}} (MSF) \cite{Pecora1998} and the negative valued regions of MSF becomes a necessary condition for occurrence of synchronization. Recently, induced synchronization has been observed in coupled chaotic systems by Schr{\"o}der {\emph{et al.}} using the method of {\emph{transient uncoupling}} \cite{Schroder2015}. This method induces synchronization in coupled systems and enhances the stability of synchronization to greater values of coupling strength \cite{Schroder2016,Aditya2016,Ghosh2018}. Further, the effect of the size of the chaotic attractors with different Lyapunov dimension in enhancing synchronization stability is studied \cite{Sivaganesh2019}. However, the direction dependence of the method transient uncoupling in clipping the phase space of chaotic attractors of the response system in a {\emph{drive-response}} scenario is yet to be studied. This paper introduces a new approach to study the direction dependence of transient uncoupling i.e., {\emph{optimal uncoupling}} and for the identification of optimal directions of implementing clipping widths to achieve greater stable synchronization in coupled chaotic systems. The following are discussed in this article. The method of transient and optimal uncoupling are briefly discussed in Section \ref{sec:2} and in Section \ref{sec:3}, the implementation of the method of optimal uncoupling in enhancing synchronization in coupled chaotic systems is presented.

\section{Transient and Optimal Uncoupling}
\label{sec:2}

The method of {\emph{transient uncoupling}} involves the clipping of phase space of the response system over the coordinate axis through the drive and response systems are unidirectionally coupled. The state equations of a $d$-dimensional chaotic system subjected to transient uncoupling driven by an identical chaotic drive system can be written as
\begin{equation}
{\bf{\dot x_2}} =   {\bf{F(x_2)}} + \epsilon \chi_{A} (\bf{x_2}) \bf{G} \times (\bf{x_1} - \bf{x_2})
\label{eqn:1}
\end{equation}
where, $\epsilon,~\chi_A$ represent the coupling strength and transient uncoupling factor and {\bf{G}} is the coupling matrix. The terms ${\bf{x_{1},x_{2}}}$ represents the state vectors of the drive and response systems and the transient uncoupling factor $\chi_A$ representing the region of phase space $A$ where, $A \subseteq \mathbb{R}^d$, is written as
\begin{equation}
\chi_A =
\begin{cases}
1 & \text{if ${\bf{x_2}} \in A$}\\
0 & \text{if ${\bf{x_2}} \notin A$}
\end{cases}
\label{eqn:2}
\end{equation}
The phase space of the response system is clipped normal to the axis of the coordinate variable $({\bf{x}}_2)_i$ where, $i=1,2,...d$, that couples the drive and response systems with respect to a point $({\bf{x}}_{2}^*)_i$ to a width $\Delta$. The clipped region of phase space is given as
 \begin{equation}
A_{\Delta} = \{ {\bf{x}}_2 \in \mathbb{R}^d : |({\bf{x}}_2)_i - ({\bf{x}}_{2}^*)_i| \le \Delta \}
\label{eqn:3}
\end{equation}
However, the clipping of phase space of the response system has not to be always restricted to any one of the coordinate axis and it can have orientations $(\theta)$ with respect to the coordinate axes. The method of finding the optimal direction $(\theta^{*})$ of applying clipping width to obtain stable synchronization leads to the evaluation of the effectiveness of clipping fraction for which synchronization is observed in the coupled systems for a fixed value of coupling strength \cite{Schroder2015} and is given as
\begin{equation}
S(\theta) = \int_0^1 s(f,\theta) df,
\label{eqn:4}
\end{equation}
where, the synchrony indicator $s(f,\theta)$ is
\begin{equation}
s(f,\theta) =
\begin{cases}
1 & \text{if $\lambda^{\perp}_{max} < 0$}\\
0 & \text{if $\lambda^{\perp}_{max} \ge 0$}
\end{cases}
\label{eqn:5}
\end{equation}
with $f$ being the temporal clipping fraction given as
\begin{equation}
f = \lim_{T \to \infty} \frac{1}{T} \int_0^T \chi_A ({\bf{x_2}}(t)) dt,
\label{eqn:6}
\end{equation}
%
The {\emph{master stability function}} being the largest transverse Lyapunov exponent $\lambda^{\perp}_{max}$ is obtained to identify the stability of synchronized states in coupled chaotic systems \cite{Pecora1998,Pecora1997}.

\section{Results and Discussion}
\label{sec:3}
We present in this section, the effect of optimal uncoupling in enhancing the stability of synchronized states and explain the novel approach in identifying the optimal directions for applying the clipping widths. The orientation of the clipping width ($\theta$) in the phase space of the  response system is considered to vary in the anticlockwise direction with respect to the $x$ or $z$-axis of the corresponding attractor discussed. The clipping width $\Delta$ oriented at an angle $\theta$ radians, has components along both the coordinate axis. Hence, the region of phase space clipped must include clipping along both the axis which indirectly implies that both of the state variables representing the attractor in the phase space must be coupled to the respective variables of the drive system. The {\emph{R{\"o}ssler}} and the {\emph{Chua's circuit}} systems are studied in this paper using this new approach to identify the optimal directions of implementing clipping widths.

\subsection{R{\"o}ssler system}
\label{sec:2.3}

The state equations of coupled {\emph{R{\"o}ssler}} systems \cite{Rossler1976} with the clipping of phase space along a particular direction can be written as
\begin{subequations}
\begin{eqnarray}
\dot x_1  &=&  -y_1 - z_1, \\ 
\dot y_1  &=&  x_1 + a y_1,\\ 
\dot z_1  &=&  b + (x_1 - c) z_1,\\
\dot x_2  &=&  -y_2 - z_2 + \epsilon \chi_A (x_1 - x_2), \\ 
\dot y_2  &=&  x_2 + a y_2+ \epsilon \chi_A (y_1 - y_2),\\ 
\dot z_2  &=&  b + (x_2 - c) z_2,
\end{eqnarray}
\label{eqn:7}
\end{subequations}
where $x_{1,2},y_{1,2},z_{1,2}$ represents the state variables of the drive and response systems. Considering the deviation of the attractor along the $z$-axis is minimum \cite{Schroder2015}, the synchronization stability of the coupled {\emph{R{\"o}ssler}} systems corresponding to the orientation of the clipping widths in the $x-y$ plane can be explored. Hence, an orientation of the clipping width in the $x-y$ plane must have its components along the corresponding coordinates axes leading to the coupling of the systems through the $x$ and $y$ state variables. Eqs. \ref{eqn:7}(d) and \ref{eqn:7}(e) indicates that for clipping widths oriented with respect to the coordinate axes of the state space vectors ($x$ and $y$) i.e., for $\theta \neq 0, \pi$ and $\theta \neq \pi/2, 3\pi/2$, the systems are unidirectionally coupled through both the $x$ and $y$ variables by the factor $\epsilon \chi_A$. For clipping widths with orientations given by $\theta = 0, \pi$ or $\theta = \pi/2, 3\pi/2$, the systems are coupled through the state variables $x$ or $y$, respectively. \\

\begin{figure}
\begin{center}
\includegraphics[width=1\textwidth]{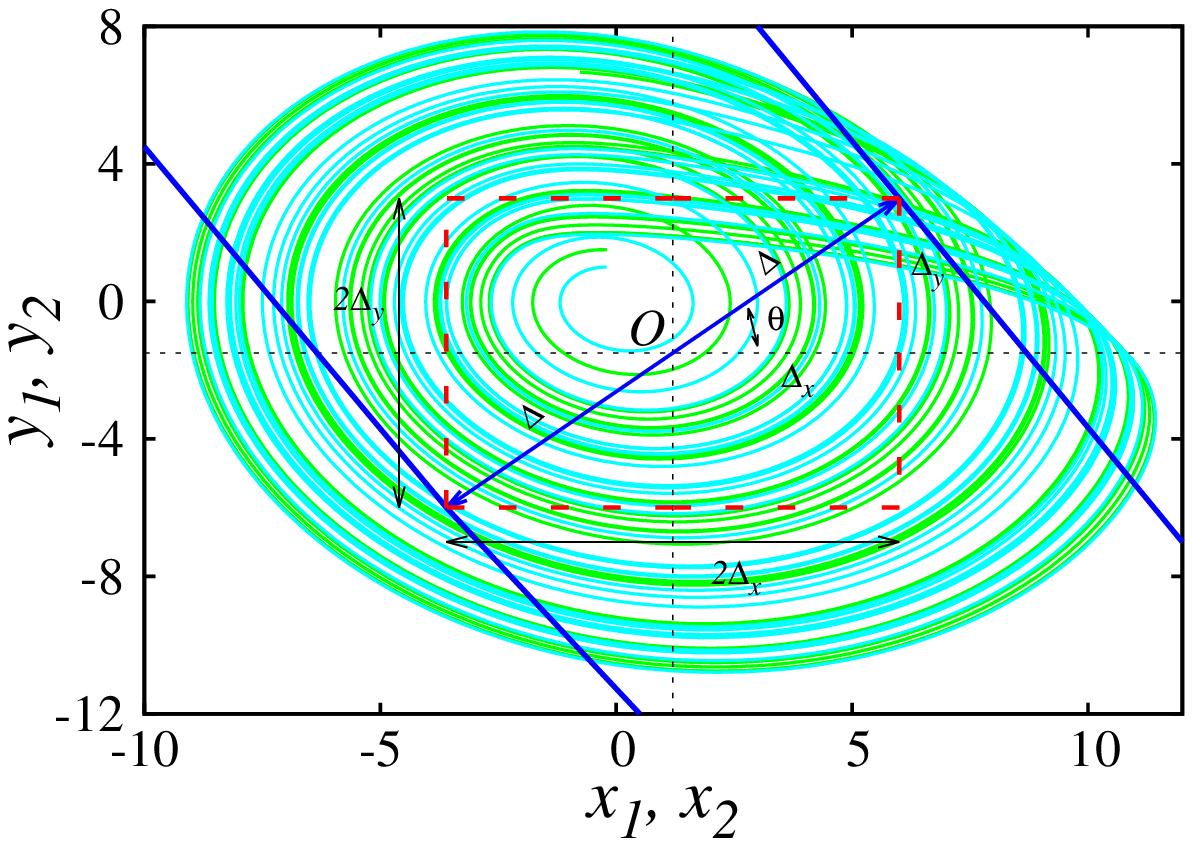}
\caption{(Color Online) {\emph{R{\"o}ssler}} system: Clipping of phase space of the chaotic attractors corresponding to the response system (cyan) along with the attractor of the drive (green) in the $x_{1,2}-y_{1,2}$ phase planes obtained for the parameters $a=0.2,b=0.2,c=5.7$ and $\epsilon = 0$. $O(x^{*},y^{*})$ is the center of the chaotic attractor with $(x^{*},y^{*})$ = (1.2,-1.5) and $OA=\Delta$ is the clipping width. Clipping is implemented along a particular direction '$\theta$' from the $x-$coordinate axis in the $(x_{2}-y_{2})$ plane over a width of $2\Delta$. $\Delta_x=\Delta cos (\theta)$ and $\Delta_y=\Delta sin (\theta)$ represent the components of the vector $\Delta$ along the coordinate axes. The coupling strength is active only over the region of phase space of the response system within the red colored box indicating the intersection of the components $\Delta_x$ and $\Delta_y$ of the vector $\Delta$ along the corresponding axis.}
\label{fig:1}
\end{center}
\end{figure}
\begin{figure}
\begin{center}
\includegraphics[width=1\textwidth]{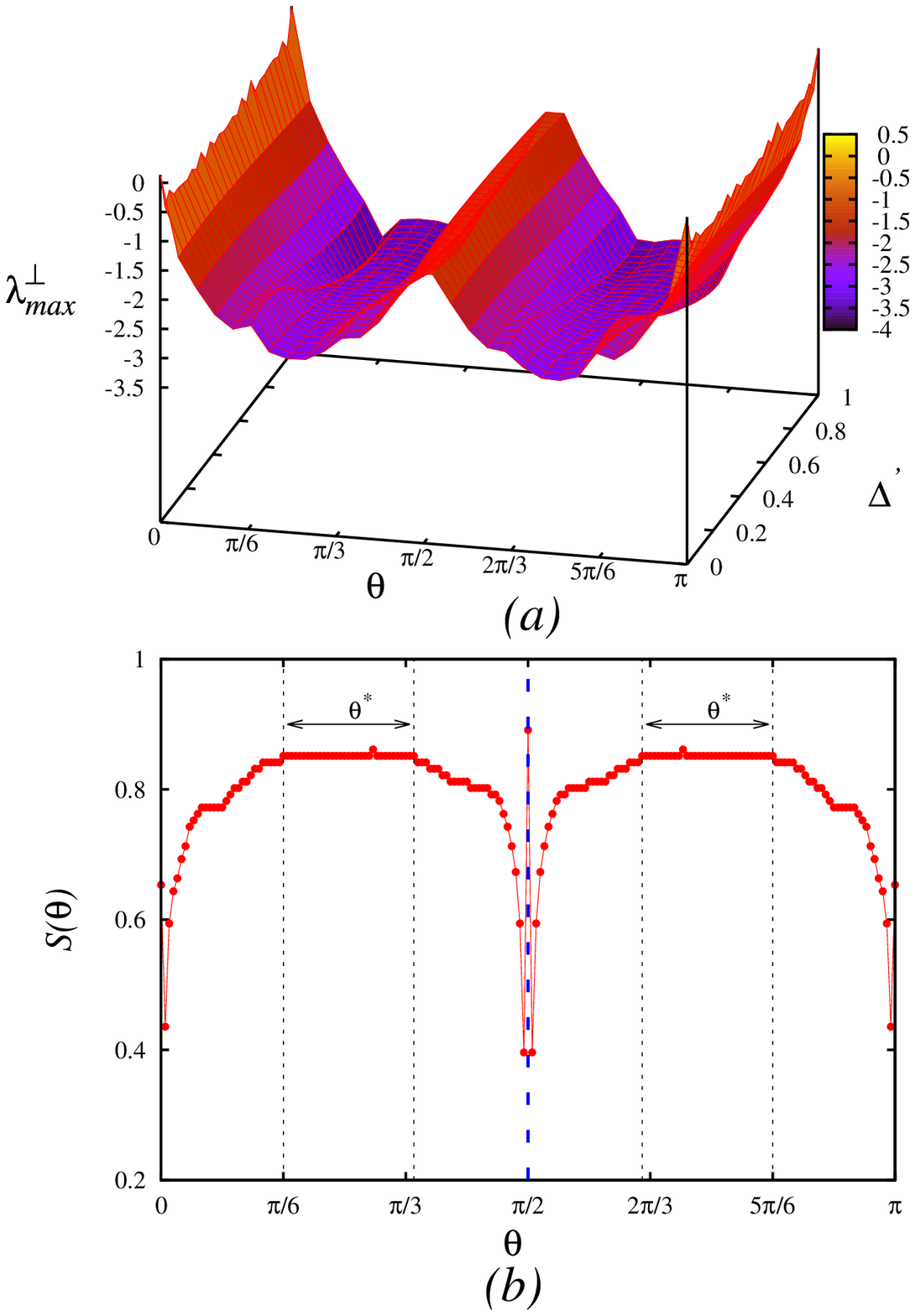}
\caption{(Color Online) {\emph{R{\"o}ssler}} system: (a) 3D plot indicating the variation of $\lambda^{\perp}_{max}$ with the orientation of clipping width $\theta$ and clipping fraction $\Delta^{'}$; (b) Variation of the effectiveness of clipping fraction $S(\theta)$ with orientation of clipping width $\theta$ indicating symmetry of the curve about the angle $\theta=\pi/2$. The optimal directions of clipping exists in the range $0.1667\pi \le \theta^{*} \le 0.3444\pi$ and $0.6556\pi \le \theta^{*} \le 0.8333\pi$, respectively.}
\label{fig:2}
\end{center}
\end{figure}
\begin{figure}
\begin{center}
\includegraphics[width=1\textwidth]{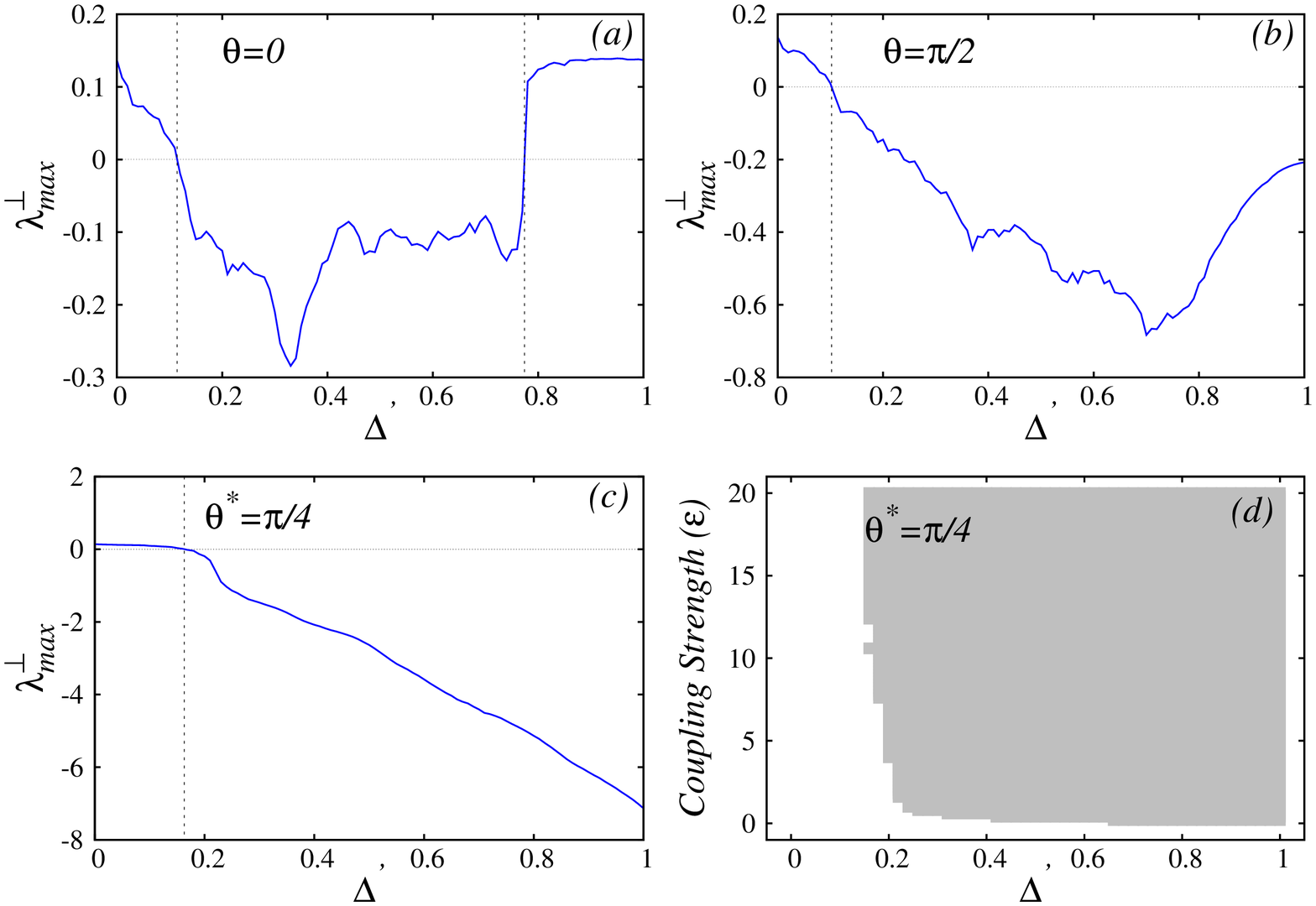}
\caption{(Color Online) {\emph{R{\"o}ssler}} system: Variation of $\lambda^{\perp}_{max}$ with clipping fraction ($\Delta^{'}$) for different orientations of the clipping width $\theta$ with the couping strength fixed at $\epsilon=10$. Variation of $\lambda^{\perp}_{max}$ with $\Delta^{'}$ for (a) $\theta=0$ i.e. $x$-coupling, indicates stable synchronization in the range $0.1445 \le \Delta^{'} \le 0.7739$; (b) $\theta=\pi/2$ i.e. $y$-coupling, indicates stable synchronization for $\Delta^{'} \ge 0.1023$; (c) Optimal direction of clipping width $\theta^{*}=\pi/4$ indicates stable synchronization for $\Delta^{'} \ge 0.163$; (d) Parameter regions for stable synchronization (gray colored) in the $\Delta^{'}-\epsilon$ plane.}
\label{fig:3}
\end{center}
\end{figure}
\begin{figure}
\begin{center}
\includegraphics[width=1\textwidth]{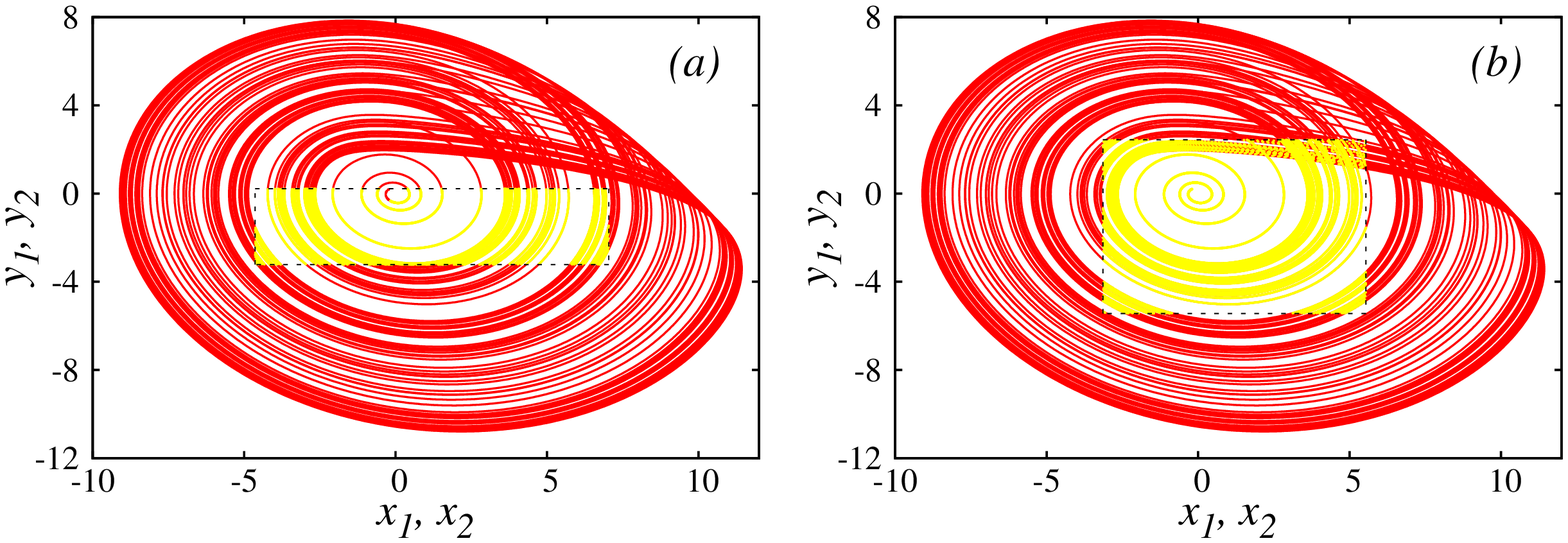}
\caption{(Color Online) {\emph{R{\"o}ssler}} system: Phase-portraits of the drive (red) and response (yellow) systems in $(x_{1,2}-y_{1,2})$ planes indicating synchronization of the coupled systems within the clipped region (black dotted box) for the parameters (a) $\theta=0.1 \pi,~\epsilon = 10,~\Delta^{'}=0.6$ and (b) $\theta=0.25 \pi,~\epsilon = 10,~\Delta^{'}=0.6$, respectively.}
\label{fig:4}
\end{center}
\end{figure}

Figure \ref{fig:1} shows the implementation of the above said method of the orientation of clipping width in the phase space of the response system along a particular direction $\theta$. The point $O(x^{*},y^{*})$ is the center of the chaotic attractor and clipping of phase space is performed to a width of $\Delta~ (OA=\Delta)$ on both sides of the center leading to a total width of $2\Delta$. The component of the clipping width $\Delta$ along the $x$ and $y$ axes are represented as $\Delta_x$ and $\Delta_y$. The intersection of the components $\Delta_x$ and $\Delta_y$ on both sides of the coordinate axes gives rise to a region of phase space $2 \Delta_x . 2 \Delta_y$ of the response system, as indicated by the red colored dotted box in Fig. \ref{fig:1}, within which the coupling of the systems is valid. This region of phase space varies with the clipping width ($\Delta$) and its orientation ($\theta$) along the $x$-coordinate axis. For $\theta=0$ and $\theta=\pi$, the coupling between the systems is only through the $x$-variable and for $\theta=\pi/2$ and $\theta=3\pi/2$, the coupling is only through the $y$-variable.  The fraction of phase space clipped along any orientation from the $x$-axis is given by the clipping fraction $\Delta^{'}=2\Delta_{x,y}/\Omega_{x,y}$ where, $\Omega_x$ and $\Omega_y$ represent the width of the chaotic attractor along the $x$ and $y$ axes, respectively. The chaotic attractor shown in Fig. \ref{fig:1} is obtained for the system parameters $a=0.2,~b=0.2,~c=5.7$ and has a Lyapunov dimension $L_d = 2.0139$.\\

The functional work steps involved in identifying the optimal directions of applying clipping widths to achieve stable synchronization is summarized as follows:

\begin{enumerate}

\item Fix the center of the chaotic attractor of the response system $({x^{*}_2}, {y^{*}_2})$ along the co-ordinate axis of the state variable coupled to the drive system.

\item For a fixed value of $\theta$ and clipping width $\Delta$ identify the region of phase space within which the coupling strength is active by resolving the horizontal $(\Delta_{x} = \Delta~ cos \theta)$ and vertical component $(\Delta_{y} = \Delta ~sin \theta)$ of the vector $OA~(OA=\Delta)$ and estimate $\lambda^{\perp}_{max}$.

\item Vary the clipping fraction $\Delta^{'}$ $(\Delta^{'}_x = 2 \Delta_x / \Omega_x,~\Delta^{'}_y = 2 \Delta_y / \Omega_y)$ in the range $0 \le \Delta^{'} \le 1$ in steps, identify the active phase-phase region of coupling strength to estimate $\lambda^{\perp}_{max}$ in each step and evaluate the effectiveness of clipping fraction $S(\theta)$ using the synchrony indicator $s(\theta)$ obtained for each step of clipping fraction.

\item Evaluate $S(\theta)$ for each value of $\theta$ by varying $\theta$ in steps in the range $0 \le \theta \le \pi$ by repeating steps 2 and 3. 

\item Plot $S(\theta)$ obtained for the corresponding value of $\theta$ to find the optimal directions $\theta^{*}$.

\end{enumerate}

The stability of synchronization of the coupled {\emph{R{\"o}ssler}} systems can be analyzed by observing the MSF to identify the optimal direction of implementing the clipping width with respect to a particular coordinate axis. Fig. \ref{fig:2}(a) shows the variation of MSF as functions of the orientation of clipping width ($\theta$) and the clipping fraction ($\Delta^{'}$). The orientation of clipping width is varied from 0 to $\pi$ radians with respect to the $x$-coordinate axis of the response system. The 3D plot shown in Fig. \ref{fig:2}(a) shows the existence of certain range of  optimal directions over which the coupled system confines to stable synchronized states. Fig. \ref{fig:2}(b) showing the variation of the effectiveness of clipping fraction $S(\theta)$ as a function of the orientation angle $\theta$ in the range $0 \le \theta \le \pi$ leads to some interesting results. Firstly, the curve is symmetrical on both sides about the orientation angle $\theta=\pi/2$. Hence, the clipping width can be sufficiently varied through the angle $0 \le \theta \le \pi/2$ to study the effectiveness of clipping orientation in coupled systems. Secondly, from Fig. \ref{fig:2}(b), it can be observed that the optimal direction $\theta^{*}$ over which the effectiveness of clipping fraction is observed and greater stable synchronized states is promised exists over a range of orientations of clipping widths. For the {\emph{R{\"o}ssler}} system, the optimal directions for stable synchronization is observed in the ranges $0.1667\pi \le \theta^{*} \le 0.3444\pi$ and $0.6556\pi \le \theta^{*} \le 0.8333\pi$, respectively. Figure \ref{fig:3} shows the variation of MSF ($\lambda^{\perp}_{max}$) with $\Delta^{'}$ for different orientations of the clipping width for a common value of the coupling strength $\epsilon=10$. Figure \ref{fig:3}(a) and \ref{fig:3}(b) showing the MSF variation with $\Delta^{'}$ for $\theta=0$ (x-coupling) and $\theta=\pi/2$ (y-coupling) indicates stable synchronized states in the range of clipping fractions $0.1145 \le \Delta^{'} \le 0.7739$ and $\Delta^{'} \ge 0.1023$, respectively. Figure \ref{fig:3}(c) shows the variation of $\lambda^{\perp}_{max}$ with $\Delta^{'}$ for the optimal direction $\theta^{*}=\pi/4$ indicating larger negative values of $\lambda^{\perp}_{max}$ for the region $\Delta^{'} \ge 0.163$. The parameter regions in the $\Delta^{'}-\epsilon$ plane indicating the negative valued regions of $\lambda^{\perp}_{max}$ for the optimal direction $\theta^{*}=\pi/4$ is shown in Fig. \ref{fig:3}(d). The synchronization of the response system with drive within the the clipped region of phase space obtained for certain values of $\theta$ is a shown in Fig. \ref{fig:4}. Figure \ref{fig:4}(a) and \ref{fig:4}(b) shows the phase-portraits in the $(x_{1,2}-y_{1,2})$ planes indicating the synchronization of dive (red) and response (yellow) systemswithin the clipped region of phase space (black dotted box), over which the coupling strength $\epsilon$ is active, for the parameters $\theta=0.1 \pi,~\epsilon=10,~\Delta^{'}=0.6$ and $\theta=0.25 \pi,~\epsilon=10,~\Delta^{'}=0.6$, respectively.\\

The method of optimal uncoupling presented above can be validated through its application to another chaotic system namely, the {\emph{Chua's}} circuit system.
\begin{figure}
\begin{center}
\centering
\includegraphics[width=1\textwidth]{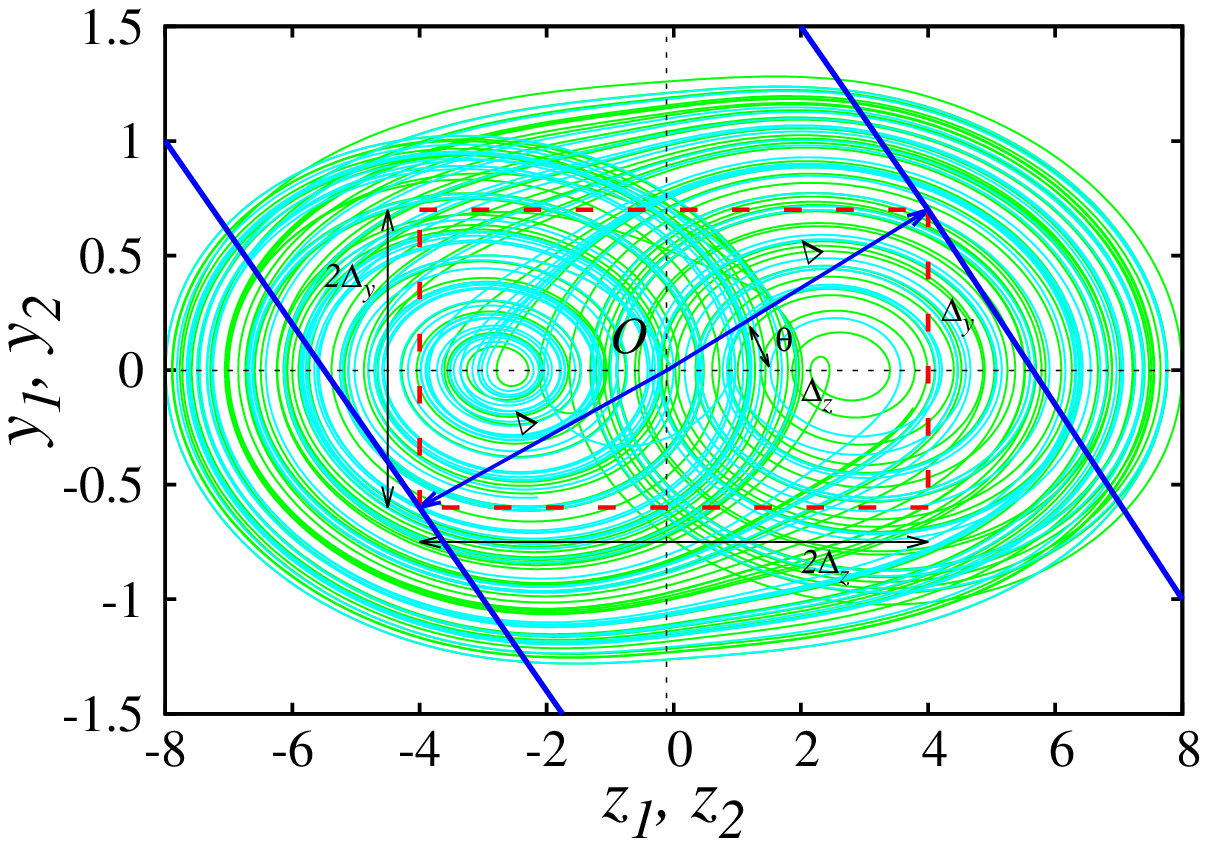}
\caption{(Color Online) {\emph{Chua's}} circuit system: Clipping of phase space of the chaotic attractors corresponding to the response system (cyan) along with the attractor of the drive (green) in the $z_{1,2}-y_{1,2}$ phase planes for $\epsilon = 0$. $O(z^{*},y^{*})$ is the center of the chaotic attractor with $(z^{*},y^{*})$ = (-0.116,~-0.002) and $OA=\Delta$ is the clipping width. Clipping is implemented along a particular direction '$\theta$' from the $z-$coordinate axis in the $(z_{2}-y_{2})$ plane over a width of $2\Delta$. $\Delta_z=\Delta~cos \theta$ and $\Delta_y=\Delta~sin \theta$ represent the components of the vector $\Delta$ along the coordinate axes. The coupling strength is active only over the region of phase space of the response system within the red colored box indicating the intersection of the components $\Delta_z$ and $\Delta_y$ of the vector $\Delta$ along the corresponding axis.}
\label{fig:5}
\end{center}
\end{figure}
\begin{figure}
\begin{center}
\centering
\includegraphics[width=1\textwidth]{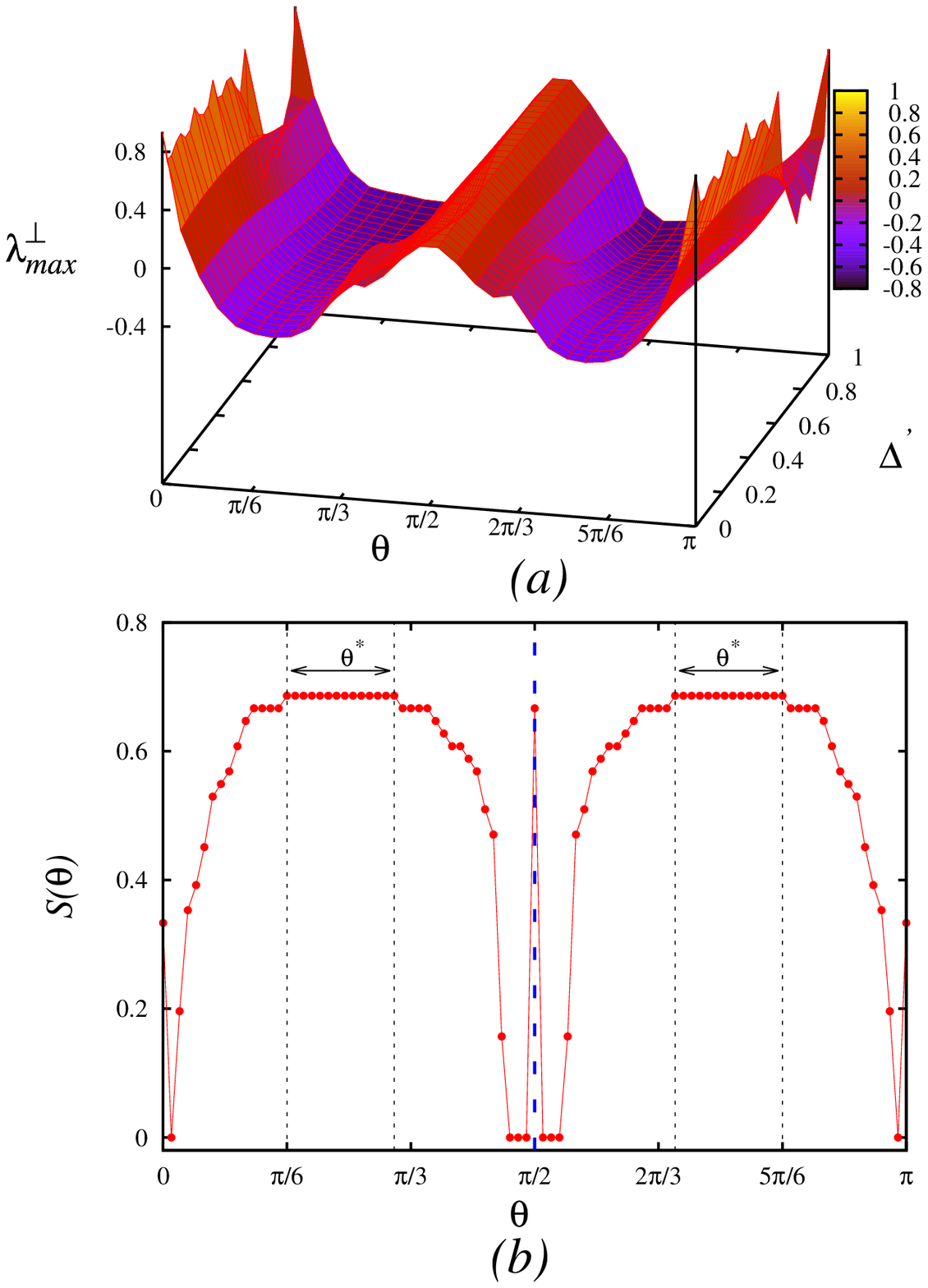}
\caption{(Color Online) {\emph{Chua's}} circuit system: (a) 3D plot indicating the variation of $\lambda^{\perp}_{max}$ with the orientation of clipping width $\theta$ and clipping fraction $\Delta^{'}$; (b) Variation of the effectiveness of clipping fraction $S(\theta)$ with orientation of clipping width $\theta$ indicating symmetry of the curve about the angle $\theta=\pi/2$. The optimal directions of clipping exists in the range $0.1667\pi \le \theta^{*} \le 0.3111\pi$ and $0.6889\pi \le \theta^{*} \le 0.8333\pi$, respectively.}
\label{fig:6}
\end{center}
\end{figure}
\begin{figure}
\begin{center}
\centering
\includegraphics[width=1\textwidth]{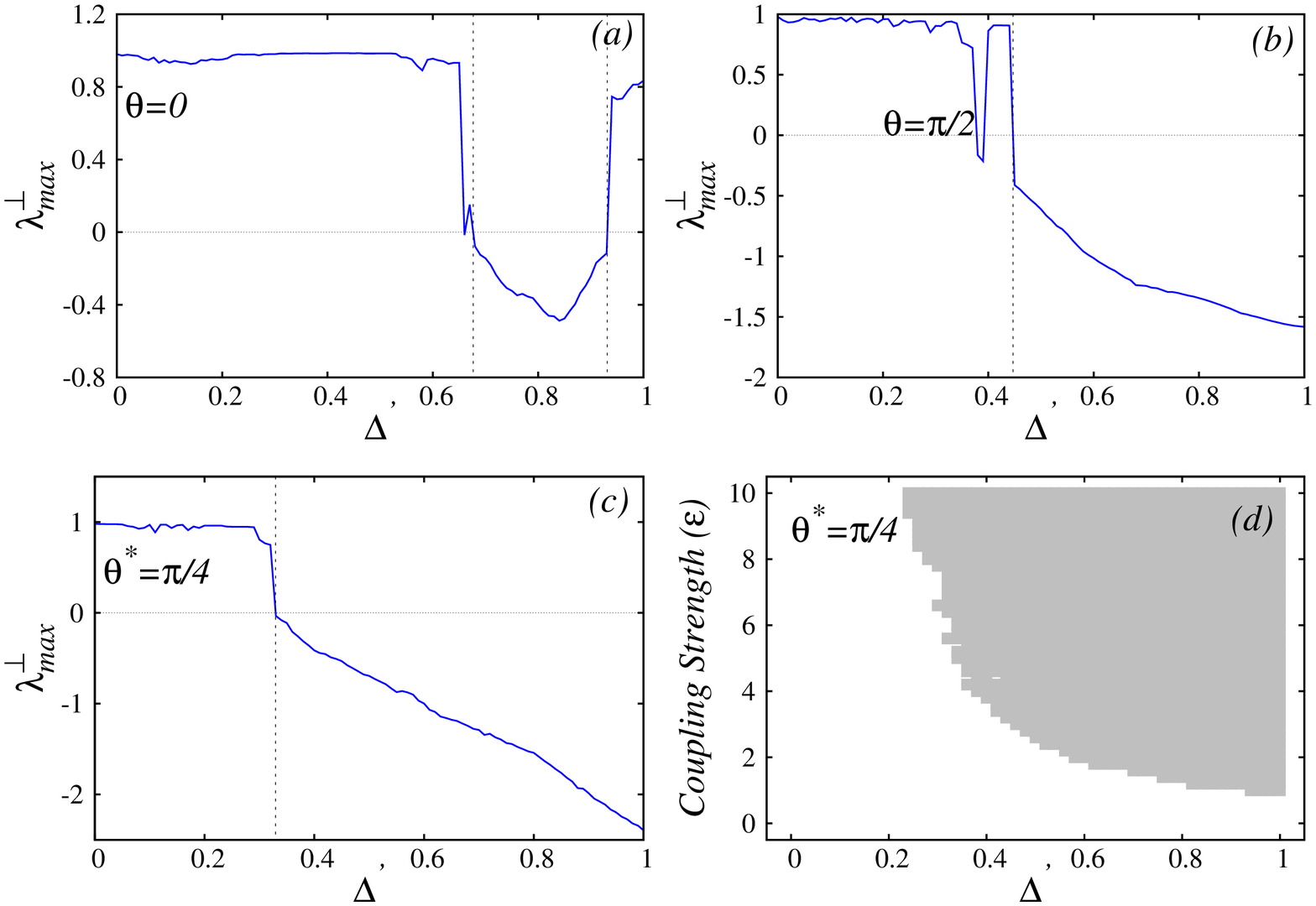}
\caption{(Color Online) {\emph{Chua's}} circuit system: Variation of $\lambda^{\perp}_{max}$ with clipping fraction ($\Delta^{'}$) for different orientations of the clipping width $\theta$ with the couping strength fixed at $\epsilon=5$. Variation of $\lambda^{\perp}_{max}$ with $\Delta^{'}$ for (a) $\theta=0$ i.e. $z$-coupling, indicates stable synchronization in the range $0.6766 \le \Delta^{'} \le 0.931$; (b) $\theta=\pi/2$ i.e. $y$-coupling, indicates stable synchronization for $\Delta^{'} \ge 0.4468$; (c) Optimal direction of clipping width $\theta^{*}=\pi/4$ indicates stable synchronization for $\Delta^{'} \ge 0.3295$; (d) Parameter regions for stable synchronization (gray colored) in the $\Delta^{'}-\epsilon$ plane.}
\label{fig:7}
\end{center}
\end{figure}
\begin{figure}
\begin{center}
\includegraphics[width=1\textwidth]{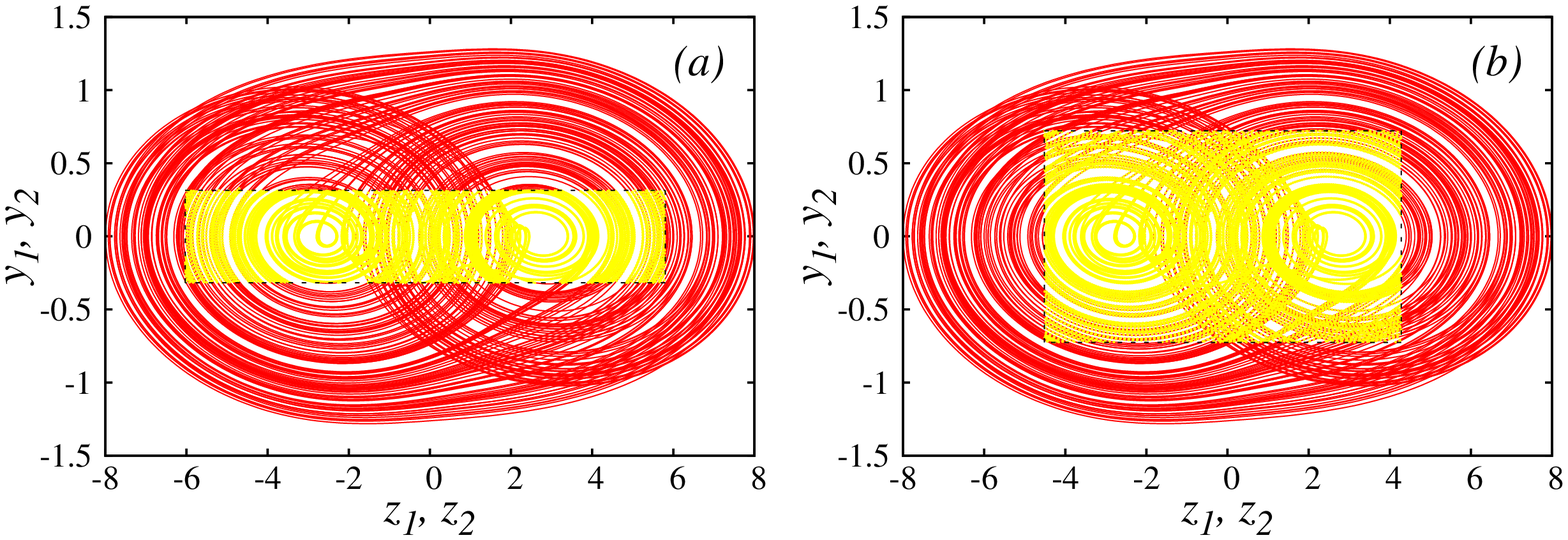}
\caption{(Color Online) {\emph{Chua's}} circuit system: Phase-portraits of the drive (red) and response (yellow) systems in $(x_{1,2}-y_{1,2})$ planes indicating synchronization of the coupled systems within the clipped region (black dotted box) for the parameters (a) $\theta=0.1 \pi,~\epsilon = 5,~\Delta^{'}=0.8$ and (b) $\theta=0.25 \pi,~\epsilon = 5,~\Delta^{'}=0.8$, respectively.}
\label{fig:8}
\end{center}
\end{figure}

\subsection{Chua's circuit system}
\label{sec:2.1}

The dynamical equations of the {\emph{Chua's circuit}} system \cite{Chua1992,Matsumoto1984,Matsumoto1985} with the drive and response systems coupled through the $y$ and $z$-variables is written as 
\begin{subequations}
\begin{eqnarray}
\dot x_1  &=&  \alpha (y_1-x_1+f(x_1)), \\ 
\dot y_1  &=&  x_1-y_1+z_1,\\ 
\dot z_1  &=&  -\beta y_1 - \gamma z_1,\\
\dot x_2  &=&  \alpha (y_2-x_2+f(x_2)), \\ 
\dot y_2  &=&  x_2-y_2+z_2 + \epsilon \chi_A (y_1 - y_2),\\ 
\dot z_2  &=&  -\beta y_2 - \gamma z_2+ \epsilon \chi_A (z_1 - z_2),
\end{eqnarray}
\label{eqn:8}
\end{subequations}
where $f(x_{1}), f(x_2)$ represent the three-segmented piecewise-linear function of the drive and response systems given as 
\begin{equation}
f(x_{1,2}) =
\begin{cases}
-bx_{1,2}+(a-b) & \text{if $x_{1,2} > 1$}\\
-ax_{1,2} & \text{if $|x_{1,2}| < 1$}\\
-bx_{1,2}-(a-b) & \text{if $x_{1,2} < -1$}
\end{cases}
\label{eqn:9}
\end{equation}
where $x_{1,2},y_{1,2},z_{1,2}$ represents the state variables of the drive and response systems. Figure \ref{fig:5} shows the implementation of clipping width for a finite orientation $\theta$ about the $z$-axis in the phase space of the response system. The point $O(z^{*},y^{*})$ is the center of the chaotic attractor and the clipping of phase space is performed to a width of $\Delta~(OA=\Delta)$ on both sides of the center leading to a total width of $2\Delta$. The intersection of the components of clipping width $\Delta_z$ and $\Delta_y$ on both sides of the coordinate axes gives rise to a region of phase space $2 \Delta_z . 2 \Delta_y$ of the response system, as indicated by the red colored dotted box in Fig. \ref{fig:5}, within which the coupling strength is valid. The clipping fraction $\Delta^{'}=2\Delta_{z,y}/\Omega_{z,y}$ where, $\Omega_z$ and $\Omega_y$ represent the width of the chaotic attractor along the $z$ and $y$ axes, respectively. The chaotic attractor shown in Fig. \ref{fig:5} is obtained for the system parameters $\alpha=10,~\beta=14.87,~\gamma=0,~a=-1.55,~b=-0.68$ and has a Lyapunov dimension $L_d = 2.1192$.\\

The stability of synchronization of the coupled {\emph{Chua}} systems can be analyzed similar to the {\emph{R{\"o}ssler}} system to identify the optimal directions of implementing the clipping width. Fig. \ref{fig:6}(a) shows the variation of MSF as functions of the orientation of clipping width ($\theta$) and the clipping fraction ($\Delta^{'}$). Fig. \ref{fig:6}(b) showing the variation of the effectiveness of clipping fraction $S(\theta)$ with orientation $\theta$ in the range $0 \le \theta \le \pi$ indicates optimal directions ($\theta^{*}$) for stable synchronization in the ranges $0.1667\pi \le \theta^{*} \le 0.3111\pi$ and $0.6889\pi \le \theta^{*} \le 0.8333\pi$, respectively. Further, the curve is symmetrical on both sides about the angle $\theta=\pi/2$. Hence, it is confirmed that optimal directions of clipping width can be obtained by studying the effectiveness of clipping fraction over the range $0 \le \theta \le \pi/2$. Figure \ref{fig:7} shows the variation of $\lambda^{\perp}_{max}$ with $\Delta^{'}$ for different orientations of the clipping width for the coupling strength $\epsilon=5$. Figures \ref{fig:7}(a) and \ref{fig:7}(b) showing the MSF variation with $\Delta^{'}$ for $\theta=0$ (z-coupling) and $\theta=\pi/2$ (y-coupling) indicates stable synchronized states in the range $0.6766 \le \Delta^{'} \le 0.931$ and $\Delta^{'} \ge 0.4468$, respectively. Figure \ref{fig:7}(c) shows the variation of $\lambda^{\perp}_{max}$ with $\Delta^{'}$ for the optimal direction $\theta^{*}=\pi/4$ indicating larger negative values of $\lambda^{\perp}_{max}$ for $\Delta^{'} \ge 0.3295$. The parameter regions in the $\Delta^{'}-\epsilon$ plane indicating the negative valued regions of $\lambda^{\perp}_{max}$ for the optimal direction $\theta^{*}=\pi/4$ is shown in Fig. \ref{fig:7}(d). The synchronization of the drive (red) and response (yellow) systems within the clipped region of phase space (black dotted box) obtained in the $z_{1,2}-y_{1,2}$ phase planes for the parameters $\theta=0.1 \pi,~\epsilon = 5,~\Delta^{'}=0.8$ and $\theta=0.25 \pi,~\epsilon = 5,~\Delta^{'}=0.8$ are as shown in Fig. \ref{fig:8}(a) and \ref{fig:8}(b), respectively. Hence, the response system synchronize with the drive within the clipped region of phase space for suitable values of $\theta$ and $\Delta^{'}$.

\section{Conclusion}

We have reported in this paper, the implementation of a novel method in enhancing the stability of synchronization observed in coupled chaotic systems through optimal uncoupling. The orientation of the clipping width in the phase space of the attractor leads to the coupling of the systems through both of the state variables representing the phase space of the attractor. The optimal directions of implementing the clipping width to achieve stable synchronization is observed over certain ranges of orientation and the functional work steps for identifying the optimal directions are presented. The method presented in the paper reveals the sufficient directions of orientation that has to be studied to identify the optimal directions which has been confirmed through studies of the {\emph{R{\"o}ssler}} and {\emph{Chua's}} circuit systems. The present study leads to the implementation of the method of transient uncoupling to enhance the synchronization stability in coupled chaotic systems over any directions in the phase space of the attractors.


\bibliography{mybibfile}

\end{document}